\documentclass[12pt,journal,compsoc]{IEEEtran}

\usepackage{graphicx}
\usepackage{epstopdf}
\usepackage{longtable}
\usepackage{caption}
\usepackage{pdflscape}
\usepackage{bbding}
\usepackage{pifont}
\usepackage{wasysym}
\usepackage{amssymb}
\usepackage[bookmarks]{hyperref}

\newcommand{\xmark}{\sffamily x}
\newcommand{\tblhead}[1]{\multicolumn{1}{| c |}{\textbf{#1}}}

\usepackage[utf8]{inputenc}

\begin{document}

\bibliographystyle{plain}

\title{The fifteen year struggle of decentralizing privacy-enhancing technology}
\author{
	Rolf~Jagerman, Wendo~Sab\'ee, Laurens~Versluis, Martijn~de~Vos,\\
	Johan~Pouwelse (course supervisor)}
\date{}

\pagestyle{empty}

\maketitle
\thispagestyle{empty}

\begin{abstract}
	Ever since the introduction of the internet, it has been void of any privacy. The majority of internet traffic currently is and always has been unencrypted. A number of anonymous communication overlay networks exist whose aim it is to provide privacy to its users. However, due to the nature of the internet, there is major difficulty in getting these networks to become both decentralized and anonymous. We list reasons for having anonymous networks, discern the problems in achieving decentralization and sum up the biggest initiatives in the field and their current status. To do so, we use one exemplary network, the Tor network. We explain how Tor works, what vulnerabilities this network currently has, and possible attacks that could be used to violate privacy and anonymity. The Tor network is used as a key comparison network in the main part of the report: a tabular overview of the major anonymous networking technologies in use today.
\end{abstract}

\section{Introduction}
	All feelings of privacy concerning browsing the internet, talking on the telephone or location tracking of cellphones are an illusion. In recent years the need for privacy-enhancing technology has become more apparent. Revelations by Edward Snowden of government misconduct and constitutional violations have sent shock waves through the internet.
	
	We failed to make an internet that is secure and private. Although much research has been done on anonymous internet communication, only few systems have been actually implemented and only one is actively being used. One of the most important factors that impact anonymity in a communication system is the number of users. A sufficiently large number of users are required for a system to make guarantees about its ability to protect the privacy of its users. This makes it difficult to introduce a new anonymous internet system, due to the initial lack of users. To support a large number of users, such a network has to be decentralized. A lack of decentralization would otherwise result in bottlenecks that place constraints on the number of users. Relying on server bandwidth donations has proven to be a difficult to sustain strategy.
	
	The most widely used anonymous communication system is Tor. In this technical report we will analyse Tor and its semi-centralized nature. Tor struggles to keep up with the bandwidth demands of its users. As the number of users increases, the need to decentralize Tor becomes more urgent. Decentralizing Tor isn't an easy task: After fifteen years of decentralization attempts, the network is still partially centralized. Only few decentralized alternatives to Tor exist, however they lack the user base to be considered safe and useful. Examples include Gnutella \cite{ripeanu2001peer}, Freenet \cite{lua2005survey} and Tapestry.
	
	This technical report is structured as following: In section \ref{sec:tor} we will give an introduction and overview of Tor. Known vulnerabilities in Tor are discussed in section \ref{sec:attacks}. After that, we will talk about decentralization and its problems in section \ref{sec:problems}. The current state of decentralized internet systems is discussed in section \ref{sec:decentralized}. A comparison of existing decentralized networks is made in section \ref{sec:comparison}. Finally, we will conclude and discuss our findings in section \ref{sec:conclusion}.
	
\section{Introduction to Tor}
	\label{sec:tor}
	
	The implementation of The Onion Router (TOR) was first described in 1996 by the U.S. Navy Research Laboratory, as a means to protect government communications from digital, as well as physical attacks, by hiding the location of the communicating party or parties \cite{goldschlag1996hiding}. The idea behind onion routing traces back to 1981, where Chaum described it in his famous paper \textquotedblleft Untraceable electronic mail, return addresses, and digital pseudonyms\textquotedblright \cite{chaum1981untraceable}.
	
	In 2002, the Tor project discontinued their old code base and re-implement the project as Tor, the Second Generation Onion Router. They introduced perfect forward secrecy, directory servers, hidden services and more \cite{dingledine2004tor}. In this section, we will explain the various components of Tor, the structure of the network, circuit creation and disadvantages of Tor.
	
	\begin{figure*}[!t]
		\centering
		\includegraphics[width=0.8\textwidth]{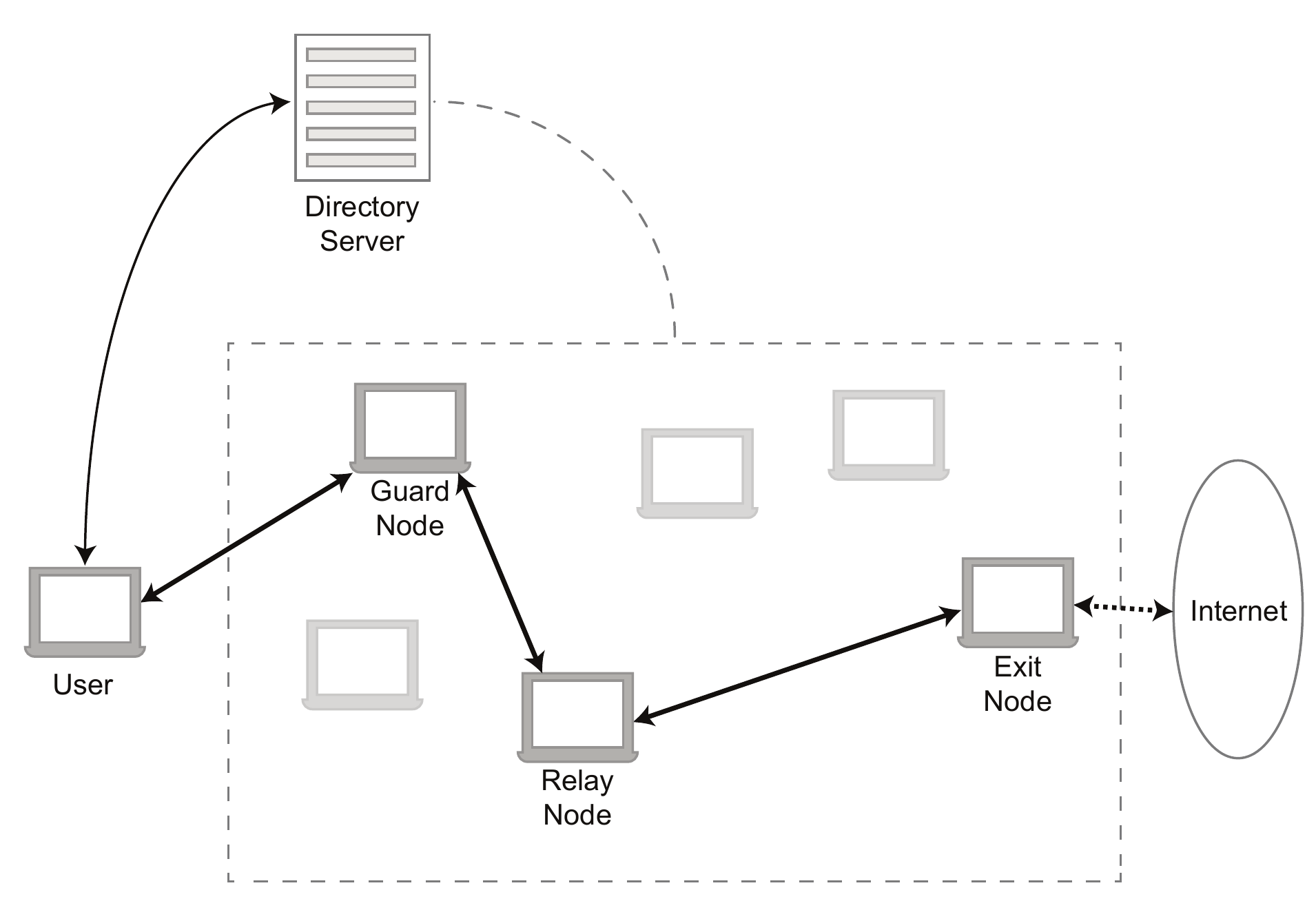}
		\caption{The components of the Tor network. After downloading the node list from the Directory Server, the user creates a circuit through a guard node, a relay node and an exit node. This circuit is used to communicate (anonymously) with the internet.}
		\label{fig:tor_layout}
	\end{figure*}

	\subsection{Onion routing}
		As described in the original design paper of The Onion Router \cite{goldschlag1996hiding}, network traffic is forwarded through a circuit of nodes, where each node only knows the previous and next node in the circuit. With a sufficiently long circuit of (independent) nodes, this means that two communicating parties can remain oblivious of each others physical location.
		
		Say we have a circuit consisting of four nodes: our trusted client node ($C$), an (entry or guard) relay node ($X$), a (middle) relay node ($Y$) and an (exit) relay node ($Z$). In this case, there are three relay nodes, but this is not necessarily always the case as more middle relay nodes can be added. A visualization of this path can be seen in figure \ref{fig:tor_layout}. Each node has its own public key and a corresponding private key. When building the circuit, our client generates a distinct secret for each of these nodes. More information about the circuit setup, can be found in \ref{ss:tor_circuit}.
		
		The payload of each packet flowing through the circuit is first encrypted with the distinct secret for last node $Z$, then with the distinct secret for node $Y$, and last with the secret for node $X$. With each layer of encryption, a header is added with the address of the next node in the circuit, plus the used distinct secret encrypted with corresponding nodes public key. 
		
		After node $X$ receives this packet from our trusted node $C$, it decrypts the attached secret with its private key, and uses that secret to decrypt the rest of the packet. The result is a header with the address for node $Y$ and the payload encrypted with secrets for the following nodes in the circuit, which is forwarded to node $Y$.
		
		As a node receives a packet from the previous node, it peels off another layer of encryption, much as how you can peel an onion layer for layer, and forwards it to the next node in the circuit (as specified in the decrypted header). When exit node $Z$ decrypts the last layer, it forwards the payload outside the network to the original destination that our trusted client $C$ tried to contact, acting as a traditional proxy.
		
		When exit node $Z$ receives a response, this whole process is applied in reverse order, encrypting the payload with its secret along the way, instead of decrypting. When our client $C$ receives the packet, it peels off all the encryption layers to retrieve the unencrypted payload.
		
		With the second generation onion routing used in Tor, a modified algorithm is used to derive the encryption keys, called telescoping path-building, which also provides perfect forward secrecy. This algorithm is described in section \ref{ss:tor_circuit}.
		
	\subsection{Directory servers}
		The original Onion Router used an unsafe, decentralized node discovery mechanism called in-band status updates. During such a status update each node broadcasts known nodes to its neighbours. An attacker could exploit this to isolate and limit the knowledge of a client, forcing connections through malicious nodes. Another disadvantage is that in-band status updates take longer to propagate throughout the network and create a global consensus.
					
		To mitigate these concerns, directory servers were introduced to Tor during its reimplementation. These directory servers keep a redundant central consensus about the network. They act as HTTP servers to which Tor nodes can publish signed information about themselves. Tor clients can in turn download this information, as seen in figure \ref{fig:tor_layout}.
					
		The information distributed by the directory servers is signed. The keys to verify these signatures are preloaded in the Tor software, along with the list of directory servers. This implies trust by the Tor client in the directory servers.

	\subsection{Relay and exit nodes}
		The Tor network consists of several components. The clients in the Tor network are known as onion proxies. The software to run an onion proxy is available for free on the Tor website \cite{torprojectwebsite} and is easy for users to configure. The onion proxies are responsible for downloading the directory information, establishing circuits across the network and handling connections from user applications.
		
		The routing in the network is done by onion routers, also called relay nodes. The relay nodes relay the data from the onion proxy to the web server across a circuit (circuits are described in \ref{ss:tor_circuit}). Each onion router is connected to every other onion router with a TLS connection \cite{tlsprotocol}. Each circuit has three type of onion routers \cite{mccoy2008shining}:
		
		\begin{itemize}
			\item{The entrance Tor router:} this router is directly connected to an onion proxy and can observe the origin of a request through the Tor network. The entrance router sends the packet to the middle Tor router.
			\item{The middle Tor router:} this router is connected to the entrance router and the exit router.
			\item{The exit Tor router:} this router is connected to the web server. Note that the exit Tor router is the only router that can observe the final destination of the request.
		\end{itemize}
		
		The first router in a circuit is the entrance router. The entrance router sends the data to one of the middle routers which forwards the data to the exit router.
			
	\subsection{Circuit creation}
		\label{ss:tor_circuit}
		
		Data on the Tor network travels over several relay nodes before it reaches its destination. Such a selection of nodes is called a circuit. To ensure both good performance and anonymity, a path is chosen using a sophisticated path selection algorithm. This algorithm selects nodes based on the bandwidth of the nodes \cite{wang2012congestion}. Nodes that have more bandwidth, have a higher probability to be chosen for the circuit creation. The same node cannot be used more than once in a single circuit.
		
		Suppose Alice is an onion proxy that wants to connect through the Tor network to a web server. Circuit creation uses the Diffie-Hellman key-exchange protocol \cite{diffiehellman} to establish a shared secret between nodes. To create a new circuit, Alice first sends a \emph{create} cell with the first half of the Diffie-Hellman handshake to the first node in her selected path (for example, $ OR_1 $). $ OR_1 $ sends a \emph{created} cell back with the second half of the key along with a hash of the final key. Now both Alice and $ OR_1 $ have a shared key they use to encrypt and decrypt data sent between them.
		
		Alice now has a connection with the first onion router in the circuit. To extend the circuit to $ OR_2 $, Alice first sends a \emph{relay extend} cell to $ OR_1 $. This cell contains the address of the next onion router in the circuit and the first half of the key to use in the communication between her and $ OR_2 $. $ OR_1 $ takes this first half of the key and sends a \emph{create} cell with this key to $ OR_2 $. When $ OR_1 $ receives a created cell, $ OR_1 $ passes this cell to Alice. Now Alice and $ OR_2 $ share a common key. The same procedure can be used to extend the circuit with more nodes.
			
	\subsection{Disadvantages}
		\label{ss:tor_disadvantages}
		
		\begin{figure*}[!t]
			\centering
			\includegraphics[width=0.95\textwidth]{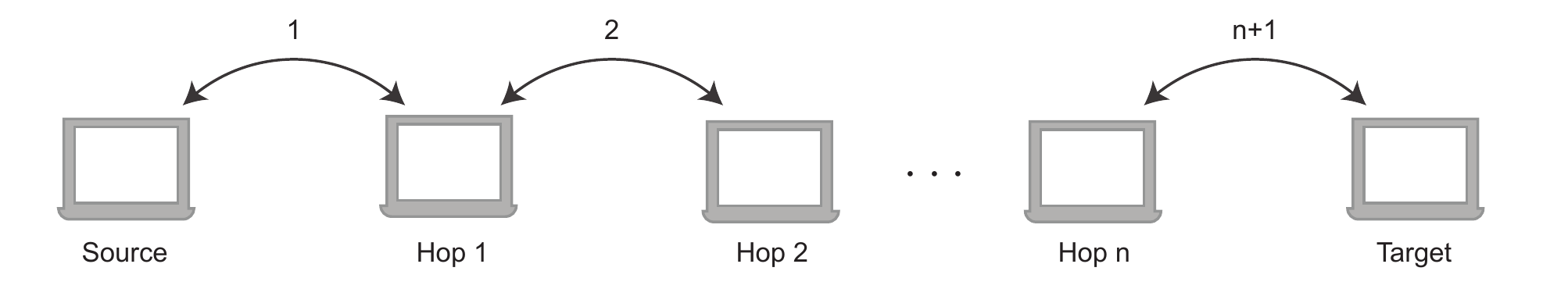}
			\caption{As traffic moves over Tor nodes, the total amount of bandwidth used in the network increases. By using $n$ hops, the total amount of network traffic would be multiplied by $2(n+1)$.}
			\label{fig:hops}
		\end{figure*}

		While Tor offers its users a high level of anonymity, there are some disadvantages using it. According to Dingledine et al \cite{dingledine2009performance}, there are six reasons why Tor is not optimal. In this section, we will summarize these reasons and explain what could be done to fix them.
		
		\begin{itemize}
			\item Tor's congestion control does not work well. The network has some problems handling bulk transfers, such as downloading large files or streaming high-quality videos. The congestion control could be improved by using an unreliable protocol such as UDP for links between Tor relays. Goldberg et al. \cite{alsabah2013pctcp} have proposed PCTCP which could improve the response time of the Tor network by 60\% and the download time of files by 30\%.	
		
			\item Some Tor users put more traffic on the network than they contribute by running an onion router. This means that these users are slowing the network down as they use more traffic than giving back. A possible solution for this is to throttle certain high-bandwidth protocols such as BitTorrent at exit nodes or at onion proxies.
		
			\item The Tor network doesn't have the capacity to handle all the users that want privacy on the internet. According to the Tor Metrics project \cite{tormetricsprojectwebsite}, it takes about 6 seconds to download 1 MiB of data.
			
			Due to the fact that traffic travels over multiple Tor nodes, the total amount of transferred data in the network multiplies. This is illustrated in figure \ref{fig:hops}. Normally, when 1 GiB is transferred over the internet, it has a network cost of 2 GiB\footnote{1 GiB upload by the sender and 1 GiB download by the receiver, or 2 GiB in total.}. By having $n$ hops, the amount of transferred traffic would be multiplied by $2(n+1)$. As Tor uses 3 hops by default, this means that a 1 GiB transfer would result in a network cost of 8 GiB. By increasing the amount of onion routing nodes in the Tor network, the capacity is increased. Incentives such as LIRA \cite{jansen13lira} could make more users run an onion router, thus increasing the capacity of the network and making it faster.	
		
			\item The current path selection algorithm of Tor doesn't distribute the load evenly over the network. The problem is that the current selection strategy is optimal when the network is fully loaded. This is not always the case. Using a better path selection algorithm could increase the capacity of the network and the overall user experience.		
		
			\item Tor clients are not optimal at handling latency and connection failures. For example, if extending a circuit fails, the entire circuit is abandoned. An improvement would be to first try to extend the circuit to some other places. If that fails, the circuit could be abandoned. Also, a better timeout mechanism could be chosen for building circuits.
			
			\item Much of the overhead of the network is in downloading the directory information. There is also overhead in the TLS connection between the nodes in the network. According to Dingledine et al, removing the empty TLS application record could reduce the overhead in the TCP/IP header by 6.3\%.	
		\end{itemize}
		
		The directory service generates overhead on the network. Replacing this central authority with a decentralized component, could reduce the overhead of the network and improve performance. In conclusion, this means that we would like to see Tor decentralized. Although much research has been done on the decentralization of Tor, it still uses centralized components today.

		\subsection{Tor stinks?}
			Tor is not only used by human rights activists. It is also used by distributors of illegal content and providers of illegal services, because it is deemed untraceable. This serious problem is disrupting Tor's public image. A recent example is the shutdown of \emph{Silk Road}, which is an online market for trading prohibited substances and other illegal goods \cite{ibtimes2013}. The market operated as a hidden Tor service. These services are accessed through an onion address and not an IP-address, hiding the physical location of the service. This makes it very hard for agencies to track and shut down these operations.

			A recent research on the content and popularity of Tor's hidden services \cite{biryukov2013content} has shown that although there are Tor hidden services that distribute illegal content, many hidden services are resources devoted to human rights, freedom of speech and information which is prohibited in some countries. It is not clear which type of service is more popular on Tor.

			Unless hidden services are used, content travels unencrypted through a Tor exit node. Users running a Tor exit node could be held responsible for distributing illegal content. The possibility of being seen as the originator of illegal content refrains users from running an exit node. A possible solution could be to filter the illegal content from the legal content. While content filtering could be a possible solution, it is an open question whether filtering is in line with the principles of Tor and the internet.
		
\section{Tor vulnerabilities and attacks}
	\label{sec:attacks}
	
	Besides the disadvantages mentioned in the previous section, Tor also suffers from several vulnerabilities that can be exploited through attacks \cite{abbott2007browser, douceur2002sybil, bauer2007low}. In this section we will summarize some of the most well known problems with Tor as well as define the following categories of attacks: browser based attacks, low-resource routing attacks, Sybil attacks and replay attacks.

	\subsection{Browser based attacks}
		Traffic analysis can be used to attack the anonymity of a user (a Tor client) browsing the web using Tor \cite{abbott2007browser}. By misusing the exit policy of Tor one can reduce the time required to perform the analysis from $O(nk)$ to $O(n+k)$ where $n$ is the number of exit nodes and $k$ is the number of entry guards.

		By running an HTTP exit node and a Tor router that eventually will act as an entry node in the network, an adversary can discover the identity of a user. The exit node injects an invisible iframe containing some JavaScript into any web page that passes through it, each sending a unique ID to a malicious web server. Every ten minutes the Tor client chooses a new circuit and eventually an unlucky Tor client picks and uses the malicious entry node that was placed in the network.

		By performing traffic analysis to compare the unique IDs of the web server and the circuits passing trough the entry node, a user can be identified. Disabling JavaScript does not mitigate this, because a similar attack can be set up only using the HTML meta refresh tag. To increase the odds of a user choosing the malicious exit node, one can run the exit node on unpopular ports. There are usually only a few exit nodes running on file sharing ports, such as 4661 to 4666. Since Tor prefers older circuits, using a denial of service attack against the older exit nodes forces Tor into creating a circuit with the malicious exit node.

		The solution for the JavaScript injection attack is disabling active content systems in the browser. For the HTML only variant one would have to use HTTPS to prevent man-in-the-middle attacks.
		
	\subsection{Low-resource routing attacks}
		Another possible attack on the anonymity of Tor is the so called low-resource routing attack \cite{bauer2007low}. With this attack, it is possible for an adversary to perform an end-to-end traffic analysis with minimal resources, thus compromising the anonymity Tor provides. The idea of this attack is that a malicious onion router can lie about its bandwidth, thus advertising a much higher bandwidth than it actually has. Because of the Tor path selection algorithm that prefers high-bandwidth nodes, the chance that a malicious entry and exit node are chosen is high.

		Once a malicious entry and exit node have been chosen, an analysis of the traffic can be done to link onion proxies with the web servers they communicate with. Experimental research in a test setting has shown that with a total of 66 non-malicious and 6 malicious nodes, is possible to compromise 46\% of the built circuits. At the request of the Tor community, this attack hasn't been tested on the live Tor network. There are some extensions and improvements to this attack. It is possible to perform a denial of service attack on well used entry nodes, forcing Tor clients to choose a new one. This improves the chance that a Tor client chooses a malicious entry node. 

		Bauer et al. proposed several solutions in their paper. The first one is to actually verify the resources of the nodes by, for example, measuring the bandwidth and/or uptime of a node. Bandwidth can be checked centralized or decentralized: the disadvantage of a centralized bandwidth check is that it generates much overhead on the network. With distributed bandwidth verification, Tor routers monitor each other but this is not enough to detect selective malicious nodes. Another solution is to restrict the amount of routers that can be on a single IP-address. The last solution proposed is to change the routing strategy.	
		
	\subsection{Sybil attacks}
		The Sybil attack is an attack where a single attacker represents itself as millions of nodes in a peer-to-peer system. Abusing this, the attacker is able to propagate false assumptions about the network to other nodes.
		
		First described by Douceur \cite{douceur2002sybil}, he mathematically proves that this attack is always possible without a central authority that certifies the participating nodes in one way or another. The exception to this rule is what he calls "extreme and unrealistic assumptions of resource parity and coordination among entities", or in other words: require all participants to do something expensive (in terms of resources) to identify themselves. This must be done within a small enough time frame, so that an attacker can't do this in sequence, but all nodes must do them in parallel.
		
		A fully distributed network that implements such a solution is the Bitcoin network \cite{nakamoto2008bitcoin}, in which computing power, and not the number of nodes is important for the general network consensus.
				
	\subsection{Replay attacks}
		A replay attack \cite{pries2008new} happens when a malicious entry node duplicates cells and sends them again. Since Tor uses the counter mode of Advanced Encryption Standard (AES-CTR) for encryption and decryption, the counter will be wrong when the duplicated package arrives causing the circuit to be destroyed.

		Using this, an accomplice exit router can, in cooperation with the entry node, discover the sender and receiver's relationship. This attack can also be used as a denial of service attack. According to Pries et al., defending against this attack is quite challenging and requires further research.

\section{Problems with decentralizing}
	\label{sec:problems}

	Decentralization is a difficult research problem. A trusted central authority simplifies bootstrapping, key management and user reputation. If one of these authorities were taken control of, anonymity could be compromised. When decentralizing a central authority, its functionality needs to be dispersed across the peers in the network. In this section we will explain the various problems involving Tor decentralization.
	
	\subsection{Incentives in decentralized systems}
		Tax evasion and environmental pollution can be seen as forms of free-riding, a phenomenon that is prominent in Tor. People predominantly use more Tor bandwidth than they donate, see section \ref{ss:tor_disadvantages}. There are multiple proposals \cite{dingledine2010building, jansen13lira} to introduce incentives into Tor, all have failed. If one would also have to build an incentive system into a decentralized system, they would have to find a way to manage the ratings of each client in the network, in such a way that they cannot be falsely modified. In other words, the reputation data has to be accurate and reliable. Besides the integrity of this data, the traffic it generates on the network should have minimal impact on the overall performance.
		
		Rahman \cite{rahman2009survey} proposes several options to build incentives in a peer-to-peer network. The first proposal that is described is the so called Warm-glow Model. This model determines the percentage of free-rides based on the probabilistic population distribution. If the percentage is above a certain threshold, the system will show signs of diminishing marginal returns.
		
		The second proposal is using monetary schemes. While it is not exemplified a lot, the main idea is to use a virtual currency as incentive. The problems with this approach are the scalability and the hidden costs of this service.
		
		The third proposal is Reciprocity-Based Schemes. Using this approach, a peer maintains a behaviour history of other peers in the network. These schemes can be based on two somewhat reciprocities: direct reciprocity or indirect reciprocity. The former are more suitable for longer relationships between peers. The latter is more scalable but they rely on third party and must handle trust issues themselves.
		
	\subsection{NAT traversal}
		A truly decentralized system requires the participating nodes to have direct connection to each other. Because of the limited availability of IPv4 network addresses, most consumer grade internet connections only provide one network address per subscriber, shared by all the devices connected to the subscribers network using Network Address Translation (NAT). With IPv4 network addresses getting more scarce, some Internet Service Providers even put more than one subscribers behind a single network address, using a carrier-grade NAT.
		
		A NAT-based system works by creating a local, private network which a NAT-enabled router connects to the internet. The local network uses network addresses from the private ranges (e.g. 10.0.0.0/8 or 192.168.0.0/16). When a local device sends a packet to the internet, the router replaces these private addresses, including the source ports, with its own public address before forwarding it to the internet. It saves these translations in a local table. Once it receives packets, it looks in this table and replaces the public network address and port with the associated private network address and port. If no corresponding entry exists in the routers table, the packet is dropped. This means that when contacting a device behind a NAT, there must be an existing entry in this table.
		
		There are several techniques to add an entry to the routers NAT table. Universal Plug and Play (UPnP) is one of those techniques, where the local device uses an HTTP request to the router to associate a port with the devices private network address. This technique is not available everywhere and sometimes considered a security risk. Different implementations of NAT require different techniques, such as hole punching, relaying or reversal, as described by Wocker et al. \cite{wacker2008nat}.

	\subsection{Bootstrapping new nodes}
		If a Tor user decides to donate some of his bandwidth by running a bridge or a relay and thus creating a new node in the network, there has to be a starting point where this new node can discover neighbours in the network to connect with. In Tor, a directory server can tell the new node what his neighbours are and where to find them \cite{dingledine2004tor}.
		
		Moving this system to a peer-to-peer base is difficult: Dingledine et al. stated that this is indeed still an open problem. With decentralized systems, there is no central directory server to tell a new node where to locate neighbours. Some systems such as Tarzan, MorphMix and Pastry \cite{rowstron2001pastry, rennhard2002introducing} are decentralized but they do suffer from performance issues.

	\subsection{Key exchange}
		With a decentralized network, using a centralized authority for managing the keys is not possible. This means that for secure communication, peers have to exchange the keys directly with each other, without a trusted party between them. Diffie-Hellman is a very popular algorithm for exchanging keys between two parties. It is used in the circuit creation in Onion Routing (see section \ref{ss:tor_circuit}) for example. However, it is possible for an adversary to manipulate the keys exchanged between two parties, making the protocol vulnerable to a man-in-the-middle attack. This weakness makes it possible for an adversary to decrypt all messages sent between the two parties.

		Tor currently uses an interactive forward-secret key-exchange protocol called the Tor Authentication Protocol (TAP) \cite{backes2012ace}. This protocol uses telescoping, which means that the initiator negotiates session keys with each successive hop in the circuit. There are several proposals for more efficient key exchange methods. One of them is ACE, an one-way authenticated key exchange protocol. The authors of this methods claim to have a 46\% efficiency improvement on the side of the client and nearly 19\% on the side of the onion routers. ACE requires clients to send one extra element in the key exchange. This does not introduce any overhead however, because the element fits in the unused space in a cell.

\section{Decentralized \mbox{privacy-enhancing} systems}
	\label{sec:decentralized}
	
	Fully decentralized systems with large scale usage are without exception based on the peer-to-peer paradigm. Many such systems have been proposed, yet only some have been implemented and are currently in use and actively maintained \cite{mislove2004ap3, rennhard2002introducing, panchenko2006nisan, rowstron2001pastry, nambiar2006salsa, freedman2002tarzan, ripeanu2001peer, androutsellis2004survey}. Here we focus on fully decentralized networks with the exception of Torsk (which is almost fully decentralized but still requires a neighbourhood authority).

	\subsection{Gnutella} 
		Gnutella is a decentralized peer-to-peer network used for distributed search of files. Since the network is fully decentralized, peers in the network are called servents, a combination of the words servers and clients. Each peer can act both as a server, answering queries, or as a client, requesting and executing search queries. In order for a client to bootstrap, a new peer connects to one of several known hosts that are almost always available \footnote{These peers can be found on http://gnutellahosts.com.}. Once the peer has joined, there are several messages a servent can send out:
	
		\begin{itemize}
			\item A peer sends a PING message to its neighbours to announce its presence. This message is forwarded to other peers and each peer sends a PONG message back.
			\item The peer can issue QUERY answers. Other peers responds with a QUERY RESPONSE message to specify whether the file that was issued in the query, was found or not.
			\item To transfer items between peers, the GET and PUSH messages are being used.
		\end{itemize}
		
		Gnutella is an unstructured network which means that the placement of data items is not based on any knowledge of the network topology nor the contents of the file. To search for a file, a flooding algorithm is used.
	
	\subsection{Freenet} 
		In Freenet, each data item is represented by a key that is independent of the location of the file. Freenet is called a loosely structured network because of this. To issue a query, the request is passed from client to client where each client makes a decision about the location to send the request next.
		
		There are three types of file keys in Freenet: the first one is called the Keyword-Signed Key (KSK) which is derived from a short description of the file. Another key is the Signed-Subspace key (SSK) which enables personal namespaces. This key contains a public and a private key. The private key is used to store the data and the public key is used in the queries for the file. The third type of key is the Content-Hash Key (CHK) which is used for updating and splitting of contents.
		
		The routing algorithm for storing and retrieving data is dynamic and can adjust to the topology of the network. Each peer only has knowledge about his neighbours. Each request has a Hops-To-Live timer which indicates how many peers the request may traverse. Each peer decrements the timer by one and when the timer reaches zero, the request isn't forwarded any more. Results of queries are being cached in intermediate nodes to reduce the time for a query response. To prevent looping of the requests, each request contains a random identifier. The peers that the request travels through, keep track of these identifiers and rejects the request if the request has already been answered by the peer.
		
	\subsection{Tapestry} 
		Tapestry is based on the Plaxton mesh data structure, which maintains pointers to nodes in the network whose IDs match the elements of a tree-like structure of ID prefixes up to a digit position. A property of Tapestry is that it offers load distribution and routing locality.
		
		Like Gnutella, peers can take the role of a client, issuing requests, and the role of a server where objects are stored. A peer can also function as a router, which forwards an incoming message. The routing algorithm is based on the destination ID of the packet. Routers are using local routing maps to route messages to the destination ID digit by digit. The routing system ensures that each peer in the system can be found in a logarithmic amount of hops.
		
		Tapestry is a fundamental component of OceanStore, a decentralized storage system. Tapestry is also used in systems such as Bayeux and SpamWatch, a decentralized spam-filtering system.

	\subsection{Pastry} 
		Pastry is very similar to Tapestry, but there are some small differences. One of these differences is the handling of network locality and data object replication. Pastry also uses the Plaxton mesh data structure for the routing algorithm. Each peer in the network gets assigned a random 128-bit identifier that is uniformly sampled from the key space. Each node can be found in about $\log(n)$ steps.

		The Pastry overlay network is used in several applications, such as Scribe, Squirrel and PAST. Scribe is a system that has been built to send multicast messages. Instead of relying on the multicast infrastructure, multicast messages are sent using only unicast services. Pastry is used to create and manage multicast groups. Scribe makes use of the organization, robustness and reliability of the Pastry network.

		Squirrel is a decentralized peer-to-peer web cache. The network uses Pastry to locate its objects and for the routing algorithm. Squirrel allows users to share its web cache with other users in the network, creating a large decentralized web cache. Squirrel however introduces some overhead when searching the cache. The challenge is to keep this overhead as low as possible.

		PAST is a large scale persistent peer-to-peer network that has been designed to store files. It is built upon the Pastry network and the main focus of PAST is providing performance, scalability and security.
			
	\subsection{MorphMix} 
		MorphMix \cite{rennhard2002introducing} functions similar to both Tor and MIX networks. It relies on nested encryption and routing traffic over multiple nodes to ensure anonymity of its users' communication. Additionally, MorphMix uses the typical behaviour of a MIX network, where it reorders messages that enter a node before sending them out.
		
		MIX networks are typically high latency and traffic in it moves slow. Messages will have to be stored on a node until enough messages have arrived to start sending them out in a random order. Often, cover traffic is used to generate enough messages for the network to obfuscate the real communication, which in turn generates a lot of bandwidth overhead. MorphMix has been developed as a low latency and high performance network. As such, it will not hold messages for very long, nor will it use cover traffic.
		
		In contrast to Tor, MorphMix does not feature a centralized node discovery mechanism. Instead, every node is free to chose a set of next nodes that will be the continuation of an anonymous tunnel. A malicious node could select a colluding node to continue the tunnel and therefore control the entire tunnel. To prevent this, a witness node is appointed to mediate the setup of an anonymous tunnel. Although this makes the possibility of such an attack more difficult, it doesn't make it impossible.
		
		An important part of MorphMix is the ability to detect malicious tunnels. Tunnels that are set up by well behaving nodes will select the next nodes in the tunnel randomly. Malicious colluding nodes, however, will specifically select nodes that are part of the malicious network to continue the tunnel. This will reveal itself in the fact that the probability of the selection of certain nodes is increased. By using this information it becomes possible, to an extent, to detect malicious tunnels.
		
	\subsection{AP3} 
		AP3 (Anonymizing Peer-to-Peer Proxy) \cite{mislove2004ap3} makes cooperative, decentralized anonymous communication possible. The AP3 system provides clients with three primitives: anonymous message delivery, anonymous channels and secure pseudonyms. Users are able to send and receive unicast, multicast and anycast messages anonymously. The strategy that AP3 is using for message delivery is similar to that of Tarzan: it relies on a network of peers to forward messages. A node along the request path, does not know whether the node from which it receives a message is the message's originator or simply another forwarding peer.
		
	\subsection{Tarzan} 
		Tarzan \cite{freedman2002tarzan} is a fully distributed peer-to-peer anonymity network. It implements a network address translator (NAT) to bridge between nodes running Tarzan and the internet. This means that services don't have to be aware of the fact they are running through Tarzan.
		
		Tarzan requires knowledge of a few existing nodes to bootstrap and uses a gossiping protocol to discover other nodes. Nambiar et al. showed however that this does not scale beyond roughly 10,000 nodes \cite{panchenko2009nisan, nambiar2006salsa}.
		
		Once Tarzan has knowledge of enough nodes, it achieves its anonymity with much like a Chaumian mix, with layered encryption and routing through multiple hops. In contrast to Tor and other networks, Tarzan uses cover traffic to provide protection against traffic analysis by a global advisory to find an initiator.
		
	\subsection{Tribler} 
		Tribler is a social-based peer-to-peer file sharing system backwards compatible with the BitTorrent protocol \cite{pouwelse2008tribler}. Tribler considers social phenomena and the sense of community as important parts of file sharing. Although the system is completely decentralized, it does not yet provide its users with anonymity. However, a modified Tor-like protocol for decentralized use is currently in beta.
		
		Tribler introduces a novel protocol called BuddyCast. Peer and content discovery use this protocol, which disseminates information epidemically. Additionally, the protocol allows users to find taste buddies, which are peers that share similar interests in files. This enables quick finding of content that a user is interested in and builds on the idea of social phenomena.
		
		Although not an anonymization network, Tribler does accomplish a lot on the topic of decentralization. The BuddyCast protocol uses BitTorrent infohashes to spread information completely decentralized throughout the network. With future work on anonymization, this can be a promising approach to anonymous file sharing.

	\subsection{NISAN} 
		NISAN, or Network Information Service for Anonymization Networks \cite{panchenko2006nisan}, is an anonymization network which implements a distributed node discovery. Not only does a central node administrator (the directory server in Tor) imply trust in those servers, Panchenko et al. also argue that a central node administration (the directory server in Tor) does not scale. The current directory server protocol was already improved two times to reduce bandwidth costs, with a fairly low amount of users.
		
		NISAN implements a DHT-based approach (Kademlia) for distributing node information, in such a way that does not require the client to know about all the nodes in the network (such as in Tor). To build a circuit, NISAN generates random IDs and searches for the closest hit throughout the network. This makes it possible to build a path with nodes picked in a random, uniform way among all nodes, without the trust of a third party.
		
		This does not fully protect against fingerprinting or bridging attacks (passive attacks), and suggest to do random walks throughout the network to mitigate that. The authors admit however this decreases the protection against an active attack.
		
	\subsection{Torsk} 
		Torsk \cite{mclachlan2009scalable} is an extension to Tor, designed to be an interoperable replacement for the circuit creation and directory servers as used by Tor. The authors argue that the current directory servers do not scale, with the percentage of the traffic in a network dedicated to node discovery growing as the number of nodes grow. With the 2009 version of Tor, they argue that 100\% of the networks traffic would consist of node discovery traffic with roughly 1.2 million clients. 
		
		Instead of the directory servers, it uses a DHT and a new neighbourhood authority. The DHT is a combination of Kademlia DHT and Myrmic DHT. Kademlia DHT was chosen because it is already widely used and it has proven itself for a large number of users.
		
		The Myrmic DHT runs on top of Kademlia, and introduces the neighbourhood authority. This authority issues certificates to nodes that participate in the DHT, but it does not participate in the DHT itself. The neighbourhood authority makes this solution not a fully decentralized one, but its role is a lot smaller than the current directory servers. This does not solve the trust issue, but it does solve the scalability issue.

	\subsection{Comparison}
		Using the properties of each of the previously described networks, we can now draw a comparison between these networks. This is done in the form of a tabular overview, see table \ref{tbl:comparison}. The networks are compared according to the following features:
	
		\begin{itemize}
			\item{Compatibility with Tor:} Is the network compatible with Tor? Could the network or some features of the network be used for the decentralization of Tor?
			\item{Public implementation:} Does a publicly available implementation exist?
			\item{Used in practice:} Is the network used in practice?
			\item{Attack resistance:} What weaknesses does the network have and which attacks are possible?
			\item{Unlinkability:} Does the network hide the identities of the sender and/or receiver?
		\end{itemize}
		
		\begin{table}[t]
			\centering
			\begin{tabular}[p]{| l | l | l | l | l | l | l | l | l |}
	\hline
	\parbox[t]{2mm}{\textbf{Name}} & \parbox[t]{2mm}{\textbf{Year}} & \parbox[t]{2mm}{\rotatebox[origin=l]{90}{\textbf{Tor interoperability}}} & \parbox[t]{2mm}{\rotatebox[origin=l]{90}{\textbf{Public implementation}}} & \parbox[t]{2mm}{\rotatebox[origin=l]{90}{\textbf{Used in practice}}} & \parbox[t]{2mm}{\rotatebox[origin=l]{90}{\textbf{(D)DoS protection}}} & \parbox[t]{2mm}{\rotatebox[origin=l]{90}{\textbf{Sybil attack protection}}} & \parbox[t]{2mm}{\rotatebox[origin=l]{90}{\textbf{MITM protection}}} & \parbox[t]{2mm}{\rotatebox[origin=l]{90}{\textbf{Unlinkability}}} \\ \hline

	Gnutella & 2000 & \xmark & \checkmark & \checkmark & \xmark & \xmark & \xmark & \xmark \\ \hline
	Freenet & 2001 & \xmark & \checkmark & \checkmark & \checkmark & \xmark & \xmark & \checkmark \\ \hline
	Tapestry & 2001 & \xmark & \checkmark & \checkmark & \checkmark & \xmark & \xmark & \xmark \\ \hline
	Pastry & 2001 & \xmark & \checkmark & \checkmark & \checkmark & \xmark & \xmark & \xmark \\ \hline
	MorphMix & 2002 & \xmark & \checkmark & \xmark & \checkmark & \xmark & ? & \checkmark \\ \hline
	AP3 & 2004 & \xmark & \xmark & \xmark & ? & ? & ? & \checkmark \\ \hline
	Tarzan & 2002 & \xmark & \xmark & \xmark & ? & \checkmark & ? & \checkmark \\ \hline
	Tribler & 2008 & \xmark & \checkmark & \checkmark & \checkmark & \xmark & \xmark & \xmark \\ \hline
	NISAN & 2009 & \xmark & \xmark & \xmark & ? & ? & \checkmark & \checkmark \\ \hline
	Torsk & 2009 & \checkmark & \xmark & \xmark & \checkmark & \checkmark & \checkmark & \checkmark \\ \hline
\end{tabular}

			\caption{A comparison of decentralized peer-to-peer overlay networks.}
			\label{tbl:comparison}
		\end{table}

\section{The documented struggle of alternative internet projects}
	\label{sec:comparison}	
	
	The largest repository of decentralization attempts is located at redecentralize.org. The aim of this repository is to \textquoteleft get decentralized products into the hands of billions\textquoteright \cite{aboutredecentralizeorg}. One of the ways they do this, is by maintaining a Github repository\footnote{Found at https://github.com/redecentralize/alternative-internet} with projects that in some way help to decentralize the internet. The struggle and pains of these projects illustrates the difficulty of decentralization. No projects succeeded in creating an alternative internet infrastructure. The projects range from self hosted cloud applications, to crypto currencies, to anonymous networks.

	%
		
	We made a significant contribution to this project index. We created a table containing each of the listed systems that shows statistics such as the total lines of code (LOC), age, number of contributors and commits, activity and (main) programming language, sourced from Ohloh. This should make it easier to filter out poorly maintained or otherwise deprecated projects. The table is available on the same Github repository.
		
	An excerpt of the table is included as table \ref{tbl:altinternet}. In this instance, we chose to sort the table on the number of commits, because we found that this most accurately represents both the maturity and activity of the projects. Other statistics such as LOC might not be very relevant on their own, because some projects include big libraries or other projects in their repository.
		
	In the table we notice that three of the entries are privacy-enhancing networks that aim to provide unlinkability (Freenet, Tor and GNUnet), plus one that is currently in the process of implementing such a feature (Tribler). Furthermore we notice that, once above a certain threshold, none of the statistics have a clear effect on a projects popularity. For example, Freenet and Tor have similar statistics, while the number of Tor users \cite{tormetricsprojectwebsite} is several orders of magnitude higher than the number of Freenet users \cite{freenetstatistics}.
		
	So although clear differences aren't directly noticeable, we hope that this comparison will provide more insight into which systems are more serious and mature than others, while also showing which ones are still actively maintained.
		
	\begin{table*}
		\centering
		\begin{tabular}[p]{| r | l | p{1.5cm} | r | r | r | r | r |}
	\hline
	\tblhead{\#} & \tblhead{Name} & \tblhead{Language} & \tblhead{Age} & \tblhead{Last activity} & \tblhead{LOC} & \tblhead{Commits} & \tblhead{Contributors} \\ \hline

	1 & ownCloud & PHP & 6 years & 2014-03-15 & 1,297 K & 34,391 & 392 \\ \hline
	2 & Freenet & Java & 13 years & 2014-03-15 & 442 K & 32,009 & 183 \\ \hline
	3 & Tor & C & 12 years & 2014-03-13 & 329 K & 29,200 & 184 \\ \hline
	4 & GNUnet & C & 8 years & 2014-03-13 & 427 K & 21,398 & 37 \\ \hline
	5 & StatusNet & PHP & 6 years & 2014-03-14 & 230 K & 14,877 & 92 \\ \hline
	6 & Diaspora* & Ruby & 3 years & 2014-03-14 & 51 K & 14,202 & 368 \\ \hline
	7 & SlapOS & Python & 8 years & 2014-03-15 & 588 K & 14,046 & 93 \\ \hline
	8 &Tahoe-LAFS & Python & 7 years & 2014-03-13 & 158 K & 11,565 & 54 \\ \hline
	9 & Tribler & Python & 8 years & 2014-03-15 & 148 K & 11,521 & 48 \\ \hline
	10 & Lorea & PHP & 6 years & 2014-03-12 & 873 K & 11,066 & 96 \\ \hline
\end{tabular}

		\caption{The top 10 projects from the Alternative Internet repository by number of commits (as of 2014-03-15).}
		\label{tbl:altinternet}
	\end{table*}
	
\section{Conclusion}
	\label{sec:conclusion}
		
	We explained how the leading privacy-enhancing technology Tor works and which components define a Tor network. Furthermore, we looked at the main disadvantages and problems Tor is currently facing. We investigated the issues around decentralization and compared systems that currently have a decentralized structure and/or mechanism.
		
	From table \ref{tbl:comparison}, we conclude that there is no fully decentralized system capable of offering Tor anonymity today. Decentralized systems that do exist such as Tarzan, I2P, Torsk or Gnutella, show promising attempts to decentralize and anonymize the internet. Yet each of these systems either lacks in performance or is vulnerable to some type of attack.
		
	For the first time we document in detail, the amount of wasted effort and pain spent in decentralization. The current generation of technology lead by Tor still has room for improvement, while the next generation is only just appearing on the horizon. The major problems involving decentralization are excruciatingly difficult to overcome. None of the projects have succeeded in making the internet secure and private.

\bibliographystyle{plain}
\bibliography{paper_references}

\begin{thebibliography}{10}

\bibitem{abbott2007browser}
Timothy~G Abbott, Katherine~J Lai, Michael~R Lieberman, and Eric~C Price.
\newblock Browser-based attacks on tor.
\newblock In {\em Privacy Enhancing Technologies}, pages 184--199. Springer,
  2007.

\bibitem{alsabah2013pctcp}
Mashael AlSabah and Ian Goldberg.
\newblock Pctcp: per-circuit tcp-over-ipsec transport for anonymous
  communication overlay networks.
\newblock In {\em Proceedings of the 2013 ACM SIGSAC conference on Computer \&
  communications security}, pages 349--360. ACM, 2013.

\bibitem{panchenko2006nisan}
Arne~Rache Andriy~Panchenko, Stefan~Richter.
\newblock In {\em NISAN: Network Information Service for Anonymization
  Networks}, 2006.

\bibitem{androutsellis2004survey}
Stephanos Androutsellis-Theotokis and Diomidis Spinellis.
\newblock A survey of peer-to-peer content distribution technologies.
\newblock {\em ACM Computing Surveys (CSUR)}, 36(4):335--371, 2004.

\bibitem{backes2012ace}
Michael Backes, Aniket Kate, and Esfandiar Mohammadi.
\newblock Ace: an efficient key-exchange protocol for onion routing.
\newblock In {\em Proceedings of the 2012 ACM workshop on Privacy in the
  electronic society}, pages 55--64. ACM, 2012.

\bibitem{bauer2007low}
Kevin Bauer, Damon McCoy, Dirk Grunwald, Tadayoshi Kohno, and Douglas Sicker.
\newblock Low-resource routing attacks against tor.
\newblock In {\em Proceedings of the 2007 ACM workshop on Privacy in electronic
  society}, pages 11--20. ACM, 2007.

\bibitem{biryukov2013content}
Alex Biryukov, Ivan Pustogarov, and Ralf-Philipp Weinmann.
\newblock Content and popularity analysis of tor hidden services.
\newblock {\em arXiv preprint arXiv:1308.6768}, 2013.

\bibitem{freenetstatistics}
Generated by~operhiem1~using pyProbe.
\newblock Freenet statistics.
\newblock Available at http://asksteved.com/stats/.

\bibitem{chaum1981untraceable}
David~L Chaum.
\newblock Untraceable electronic mail, return addresses, and digital
  pseudonyms.
\newblock {\em Communications of the ACM}, 24(2):84--90, 1981.

\bibitem{dingledine2004tor}
Roger Dingledine, Nick Mathewson, and Paul Syverson.
\newblock Tor: The second-generation onion router.
\newblock Technical report, DTIC Document, 2004.

\bibitem{dingledine2009performance}
Roger Dingledine and Steven~J Murdoch.
\newblock Performance improvements on tor or, why tor is slow and what we’re
  going to do about it.
\newblock {\em Online:
  http://www.torproject.org/press/presskit/2009-03-11-performance.pdf}, 2009.

\bibitem{dingledine2010building}
Roger Dingledine, Dan~S Wallach, et~al.
\newblock Building incentives into tor.
\newblock In {\em Financial Cryptography and Data Security}, pages 238--256.
  Springer, 2010.

\bibitem{douceur2002sybil}
John~R Douceur.
\newblock The sybil attack.
\newblock In {\em Peer-to-peer Systems}, pages 251--260. Springer, 2002.

\bibitem{freedman2002tarzan}
Michael~J Freedman and Robert Morris.
\newblock Tarzan: A peer-to-peer anonymizing network layer.
\newblock In {\em Proceedings of the 9th ACM conference on Computer and
  communications security}, pages 193--206. ACM, 2002.

\bibitem{goldschlag1996hiding}
David~M Goldschlag, Michael~G Reed, and Paul~F Syverson.
\newblock Hiding routing information.
\newblock In {\em Information Hiding}, pages 137--150. Springer, 1996.

\bibitem{diffiehellman}
IETF.
\newblock The diffie-hellman key agreement method.
\newblock Available at https://www.ietf.org/rfc/rfc2631.txt.

\bibitem{tlsprotocol}
IETF.
\newblock The transport layer security (tls) protocol.
\newblock Available at http://tools.ietf.org/html/rfc5246.

\bibitem{jansen13lira}
Rob Jansen, Aaron Johnson, and Paul Syverson.
\newblock Lira: Lightweight incentivized routing for anonymity.

\bibitem{lua2005survey}
Eng~Keong Lua, Jon Crowcroft, Marcelo Pias, Ravi Sharma, Steven Lim, et~al.
\newblock A survey and comparison of peer-to-peer overlay network schemes.
\newblock {\em IEEE Communications Surveys and Tutorials}, 7(1-4):72--93, 2005.

\bibitem{mccoy2008shining}
Damon McCoy, Kevin Bauer, Dirk Grunwald, Tadayoshi Kohno, and Douglas Sicker.
\newblock Shining light in dark places: Understanding the tor network.
\newblock In {\em Privacy Enhancing Technologies}, pages 63--76. Springer,
  2008.

\bibitem{mclachlan2009scalable}
Jon McLachlan, Andrew Tran, Nicholas Hopper, and Yongdae Kim.
\newblock Scalable onion routing with torsk.
\newblock In {\em Proceedings of the 16th ACM conference on Computer and
  communications security}, pages 590--599. ACM, 2009.

\bibitem{mislove2004ap3}
Alan Mislove, Gaurav Oberoi, Ansley Post, Charles Reis, Peter Druschel, and
  Dan~S Wallach.
\newblock Ap3: Cooperative, decentralized anonymous communication.
\newblock In {\em Proceedings of the 11th workshop on ACM SIGOPS European
  workshop}, page~30. ACM, 2004.

\bibitem{nakamoto2008bitcoin}
Satoshi Nakamoto.
\newblock Bitcoin: A peer-to-peer electronic cash system.
\newblock {\em Consulted}, 1:2012, 2008.

\bibitem{nambiar2006salsa}
Arjun Nambiar and Matthew Wright.
\newblock Salsa: a structured approach to large-scale anonymity.
\newblock In {\em Proceedings of the 13th ACM conference on Computer and
  communications security}, pages 17--26. ACM, 2006.

\bibitem{panchenko2009nisan}
Andriy Panchenko, Stefan Richter, and Arne Rache.
\newblock Nisan: network information service for anonymization networks.
\newblock In {\em Proceedings of the 16th ACM conference on Computer and
  communications security}, pages 141--150. ACM, 2009.

\bibitem{ibtimes2013}
Charles Poladian.
\newblock Silk road shut down by fbi, owner ross william ulbricht, 'dread
  pirate roberts,' arrested.
\newblock Available at
  http://www.ibtimes.com/silk-road-shut-down-fbi-owner-ross-william-ulbricht-dread-pirate-roberts-arrested-1413966.

\bibitem{pouwelse2008tribler}
Johan~A Pouwelse, Pawel Garbacki, Jun Wang, Arno Bakker, Jie Yang, Alexandru
  Iosup, Dick~HJ Epema, Marcel Reinders, Maarten~R Van~Steen, and Henk~J Sips.
\newblock Tribler: a social-based peer-to-peer system.
\newblock {\em Concurrency and Computation: Practice and Experience},
  20(2):127--138, 2008.

\bibitem{pries2008new}
Ryan Pries, Wei Yu, Xinwen Fu, and Wei Zhao.
\newblock A new replay attack against anonymous communication networks.
\newblock In {\em Communications, 2008. ICC'08. IEEE International Conference
  on}, pages 1578--1582. IEEE, 2008.

\bibitem{aboutredecentralizeorg}
Redecentralize Project.
\newblock About redecentralize.org.
\newblock Available at http://redecentralize.org/about/.

\bibitem{rahman2009survey}
Muntasir~Raihan Rahman.
\newblock A survey of incentive mechanisms in peer-to-peer systems, 2009.

\bibitem{rennhard2002introducing}
Marc Rennhard and Bernhard Plattner.
\newblock Introducing morphmix: peer-to-peer based anonymous internet usage
  with collusion detection.
\newblock In {\em Proceedings of the 2002 ACM workshop on Privacy in the
  Electronic Society}, pages 91--102. ACM, 2002.

\bibitem{ripeanu2001peer}
Matei Ripeanu.
\newblock Peer-to-peer architecture case study: Gnutella network.
\newblock In {\em Peer-to-Peer Computing, 2001. Proceedings. First
  International Conference on}, pages 99--100. IEEE, 2001.

\bibitem{rowstron2001pastry}
Antony Rowstron and Peter Druschel.
\newblock Pastry: Scalable, decentralized object location, and routing for
  large-scale peer-to-peer systems.
\newblock In {\em Middleware 2001}, pages 329--350. Springer, 2001.

\bibitem{wacker2008nat}
Arno Wacker, Gregor Schiele, Sebastian Holzapfel, and Torben Weis.
\newblock A nat traversal mechanism for peer-to-peer networks.
\newblock In {\em Peer-to-Peer Computing}, pages 81--83, 2008.

\bibitem{wang2012congestion}
Tao Wang, Kevin Bauer, Clara Forero, and Ian Goldberg.
\newblock Congestion-aware path selection for tor.
\newblock In {\em Financial Cryptography and Data Security}, pages 98--113.
  Springer, 2012.

\bibitem{tormetricsprojectwebsite}
Tor Metrics~Project website.
\newblock Tor project: Anonimity online.
\newblock Available at https://metrics.torproject.org.

\bibitem{torprojectwebsite}
Tor~Project website.
\newblock Tor project: Anonimity online.
\newblock Available at http://torproject.org.

\end{thebibliography}

\end{document}